\documentclass[review,number,sort&compress]{elsarticle}

\usepackage{amssymb}
\usepackage{amsmath}
\usepackage{booktabs}
\usepackage{multirow}
\usepackage{lineno}

\emergencystretch=\maxdimen
\mathchardef\mhyphen="2D

\usepackage[]{caption2}

\journal{Nucl. Instrum. Methods A}

\begin{document}

\begin{frontmatter}



\title{Multi-layer plastic scintillation detector for intermediate- and high-energy neutrons with $\it{n}$-$\gamma$ discrimination capability}

\author[buaa]{L.~Yu\corref{cor1}}
\cortext[cor1]{Corresponding author.}
\ead{yuleibuaa@buaa.edu.cn}
\author[buaa]{S.~Terashima\corref{cor2}}
\cortext[cor2]{Corresponding author.}
\ead{tera@buaa.edu.cn}
\author[rcnp]{H.J.~Ong}
\author[rcnp]{P.Y.~Chan}
\author[buaa,rcnp]{I.~Tanihata}
\author[rcnp]{C.~Iwamoto}
\author[rcnp,vietnam]{D.T.~Tran}
\author[rcnp]{A.~Tamii}
\author[rcnp]{N.~Aoi}
\author[kyoto]{H.~Fujioka}
\author[rcnp]{G.~Gey}
\author[rcnp]{H.~Sakaguchi}
\author[kyoto]{A.~Sakaue}
\author[buaa]{B.H.~Sun}
\author[rcnp]{T.L.~Tang}
\author[buaa]{T.F.~Wang}
\author[tokyo]{Y.N.~Watanabe}
\author[buaa]{G.X.~Zhang}

\address[buaa]{\it School of Physics and Nuclear Energy Engineering, Beihang University, Beijing 100191, China}
\address[rcnp]{\it Research Center for Nuclear Physics (RCNP), Osaka University, Ibaraki, Osaka 567-0047, Japan}
\address[vietnam]{\it Institute of Physics, Vietnam Academy of Science and Technology, Hanoi 100000, Vietnam}
\address[kyoto]{\it Department of Physics, Kyoto University, Kyoto 606-8502, Japan}
\address[tokyo]{\it Department of Physics, University of Tokyo, Tokyo 113-0033, Japan}

\begin{abstract}
A new type of neutron detector, named Stack Structure Solid organic Scintillator (S$^4$), consisting of multi-layer plastic scintillators with capability to suppress low-energy $\gamma$ rays under high-counting rate has been constructed and tested. To achieve $\it{n}$-$\gamma$ discrimination, we exploit the difference in the ranges of the secondary charged particles produced by the interactions of neutrons and $\gamma$ rays in the scintillator material. The thickness of a plastic scintillator layer was determined based on the results of Monte Carlo simulations using the Geant4 toolkit. With layer thicknesses of 5 mm, we have achieved a good separation between neutrons and $\gamma$ rays at 5 MeV$_{\rm ee}$ threshold setting. We have also determined the detection efficiencies using monoenergetic neutrons at two energies produced by the $\it{d}$+$\it{d}\to\it{n}$+$^{3}$He reaction. The results agree well with the Geant4 simulations implementing the Li$\grave{\rm e}$ge Intranuclear Cascade hadronic model (INCL++) and the high-precision model of low-energy neutron interactions (NeutronHP).
\end{abstract}


\end{frontmatter}

\section{Introduction}
\label{1}
Detection of intermediate- and high-energy neutrons is important to identify reaction channels and to extract nuclear structure information for experiments using nuclear reactions, e.g. ($p$,$ppn$) \cite{p2pn}, ($p$,$nd$) \cite{pnd} and ($e$,$e'pn$) \cite{eepn1,eepn2} reactions. However, neutron detection in an experimental environment has remained a challenge due mainly to $\gamma$-ray background and high counting rate. The most dominant sources of background are prompt and delayed $\gamma$ rays from the reaction target, and $\gamma$ rays from the surroundings. Such background cannot be easily eliminated by shielding materials, and in some instances, may result in a high counting rate.

One of the suppression methods of prompt $\gamma$ rays from the reaction target is the time-of-flight (TOF) method since all $\gamma$ rays travel at the speed of light regardless of their energies. This method is usually adopted by neutron detectors that use large-area plastic scintillators, such as LAND \cite{LAND} at GSI and HAND \cite{SubediPhD,HANDPRL} at JLab. Although plastic scintillators can operate at a high rate, a longer flight path, which may lead to a limited solid angle, is usually necessary both to achieve a better $\it{n}$-$\gamma$ discrimination and to reduce the counting rate mainly produced near the reaction target. The TOF method, moreover, cannot eliminate the time-uncorrelated $\gamma$ rays which also exist in most experimental conditions. Such $\gamma$-ray background can be suppressed by increasing the pulse height detection threshold, but this will be at the expense of a reduced neutron detection efficiency. An alternative to separate neutrons from $\gamma$ rays is via the pulse shape discrimination ($\mbox{\rm PSD}$) technique which utilizes the difference in the slow decay components of the induced light output of organic scintillators \cite{PSD}. Neutron detectors with PSD capability are now commercially available in the form of liquid scintillators \cite{liquid}, e.g. DEMON \cite{DEMON} and Neutron Shell \cite{NeutronShell}, as well as the recently developed solid scintillator EJ-299-33 \cite{EJ299}. These scintillation detectors offer neutron detection with $\it{n}$-$\gamma$ discrimination capability at a low detection energy threshold, but may suffer from pile-up in a high-rate environment because of the longer tail components of the induced pulses compared with normal plastic scintillators. Neutron detectors employing the two conventional techniques described above are not always able to fulfill our demands for the detection of high-energy neutrons, especially in an overwhelming background environment. Therefore, a neutron detector with fast counting and $\it{n}$-$\gamma$ discrimination capabilities at a low energy threshold is highly desirable.

In this article, we report on the development of a new type of neutron detector, named Stack Structure Solid organic Scintillator (S$^4$), which has the capability to suppress low-energy $\gamma$ rays efficiently. The detector consists of plastic scintillators with short decay time, and is thus highly affordable and flexible compared with liquid scintillators. This detector offers the capability to discriminate neutrons and $\gamma$ rays via exploitation of the difference of ranges of secondary charged particles, protons and electrons typically, in a plastic scintillator. Because it does not require timing information for $\it{n}$-$\gamma$ separation, it can be placed closer to the reaction target to gain solid angle. The principle of the discrimination technique and related simulations are presented in Section \ref{2}. Sections \ref{3} and \ref{4} describe the configuration and experimental performance of the detector. A conclusion and the future prospect are given in Sections \ref{5} and \ref{6}.

\section{Principle of $\it{n}$-$\gamma$ discrimination and design concept}
\label{2}
\subsection{Principle of $\it{n}$-$\gamma$ discrimination}
\label{2.1}
\begin{figure}
\centering
\includegraphics[width=7cm,clip]{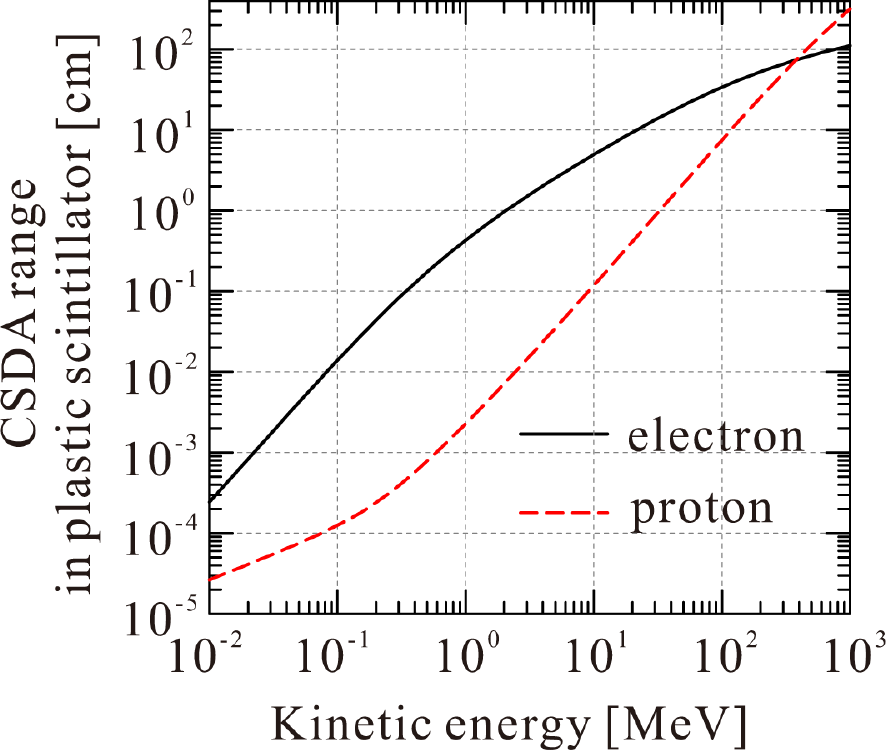}
\caption{\label{range}
Ranges of electron and proton in plastic scintillator calculated within the Continuous Slowing Down Approximation (CSDA) \cite{CSDA}. The density of plastic scintillator is 1.03 g/cm$^{3}$.}
\end{figure}

The main secondary particles generated by neutrons and $\gamma$ rays in an organic scintillator material are protons or carbon ions and electrons, respectively. Given the same energy deposit in the scintillator, the ranges of the electrons and other particles are significantly different. Figure \ref{range} shows the ranges of electron and proton in plastic scintillator calculated within the Continuous Slowing Down Approximation (CSDA) \cite{CSDA}. Below 100 MeV, the CSDA range of a proton (or a carbon ion) is about one or two orders of magnitude shorter than that of an electron. Such difference provides an important means to distinguish the neutron and $\gamma$-ray events.

\begin{figure}
\centering
\includegraphics[width=7cm,clip]{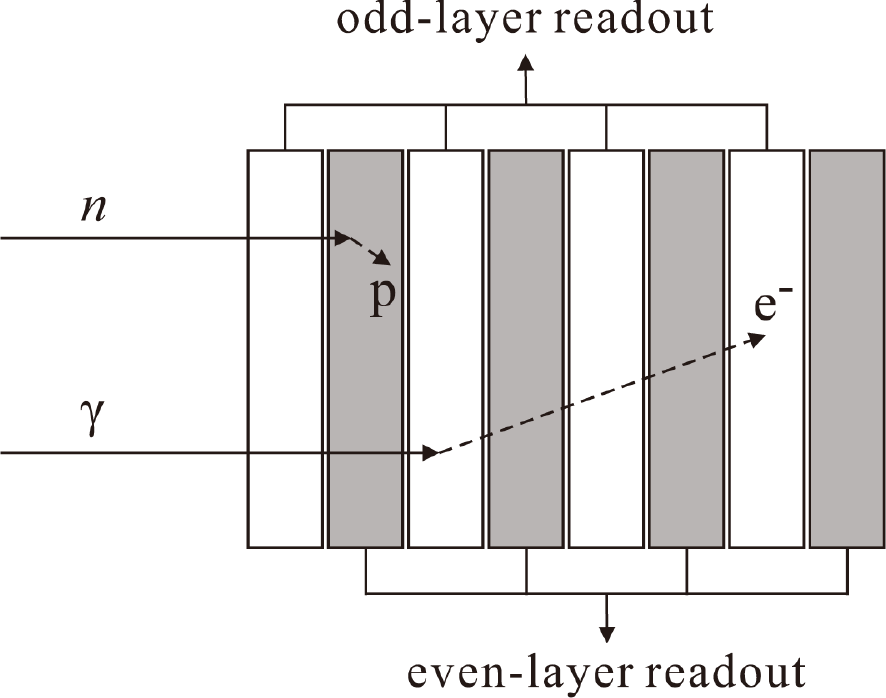}
\caption{\label{multilayer}
Illustration of the typical ranges of secondary particles induced by neutrons and $\gamma$ rays with the same energy, and the suggested readout method for the multi-layer plastic scintillators.}
\end{figure}

To this end, we consider a detector consisting of multiple layers of plastic scintillators. The most straightforward way to obtain the range of a secondary particle is to read out the signal from every single layer and identify the number of layers with signals. This method, however, requires a large number of readout electronics. As shown in Fig. \ref{range}, the recoiled protons from neutrons of energy below 100 MeV have ranges from sub-millimeter up to centimeter scale, thus most of them can be stopped in a layer of scintillator with a sub-centimeter thickness. On the other hand, the secondary electrons of the same energy have ranges from a few to more than a hundred times longer than protons and can penetrate many layers of scintillators of the same thickness. To distinguish this difference, only the information of the energy-loss difference in neighboring scintillators is necessary. The simplest way to detect this difference is to take two signals, one from the odd layers and the other from the even layers, as illustrated in Fig. \ref{multilayer}. By reading out signals from the odd and even layers, denoted as $E_{\rm odd}$ and $E_{\rm even}$, respectively, one can define the total signal $E_{\rm all}$ and the balance ratio $\rho$ as,
\begin{eqnarray}
\label{eq1}
\begin{aligned}
& E_{\rm all}=E_{\rm even}+E_{\rm odd}, \\
& \rho=(E_{\rm even}-E_{\rm odd})/(E_{\rm even}+E_{\rm odd}).
\end{aligned}
\end{eqnarray}
$\rho$ equals to $-1$ or 1 if a secondary particle stops within the layer where the conversion takes place. The $\rho$ values are expected to be around zero if a secondary particle penetrates many layers, since the deposited energy will be shared by both odd and even layers. Thus, neutron and $\gamma$ ray can be separated by $\rho$ when an appropriate thickness of plastic scintillator is selected.

\subsection{Design concept based on Monte Carlo simulations}
\label{2.2}
To investigate the feasibility of the $\it{n}$-$\gamma$ discrimination and to determine the appropriate thickness of plastic scintillator, we performed Monte Carlo simulations using the Geant4 toolkit version 10.2.p02 \cite{geant4,10.2}. We employed the conventional electromagnetic and elastic hadronic scattering models, coupled with the Li$\grave{\rm e}$ge Intranuclear Cascade model (INCL++) \cite{INCL1,INCL2} for inelastic channels above 20 MeV, and the high-precision neutron model (NeutronHP) for all hadronic processes below 20 MeV. The neutron multiple scatterings were also considered for all simulations in the present work.

For simplicity as well as practical reason, we fixed the total thickness of the plastic scintillators to 80 mm, and assumed the scintillators to be of sufficiently large area in all simulations. The simulations were performed assuming several different layer thicknesses: 1, 5 and 10 mm, which correspond to 80, 16 and 8 layers of scintillators in total, respectively. A pencil beam of generated particles was injected perpendicular to the center of the detector. We assumed incident neutrons and $\gamma$ rays with uniform energy distribution from 20 to 60, and from 0 to 10 MeV, respectively. The selected 20-60 MeV range is the typical energy range of the recoil neutrons, e.g. in the ($p$,$nd$) reaction \cite{pnd}, while 10 MeV is almost the maximum energy of $\gamma$ rays from nuclear excited states below particle-emission thresholds. Here, we generated equal number of neutrons and $\gamma$ rays per unit of energy, namely $N^{n}=4\times N^{\gamma}$ in total, where the superscript indicates the type of particle. The energy response of the plastic scintillators is expressed in terms of the electron-equivalent energy (MeV$_{\rm ee}$) using the equations taken from Ref.\cite{MeVee}, since the light outputs of the scintillator produced by an electron and a proton are different functions of the energy loss.

In the following subsections, we investigate the characteristics of the $\it{n}$-$\gamma$ discrimination qualitatively and quantitatively, and choose the appropriate layer thickness for the prototype S$^{4}$ detector for our experimental purpose.

\subsubsection{$\rho$ distributions and principle for layer-thickness determination}
\label{2.2.1}
\begin{figure}
\centering
\includegraphics[width=9cm,clip]{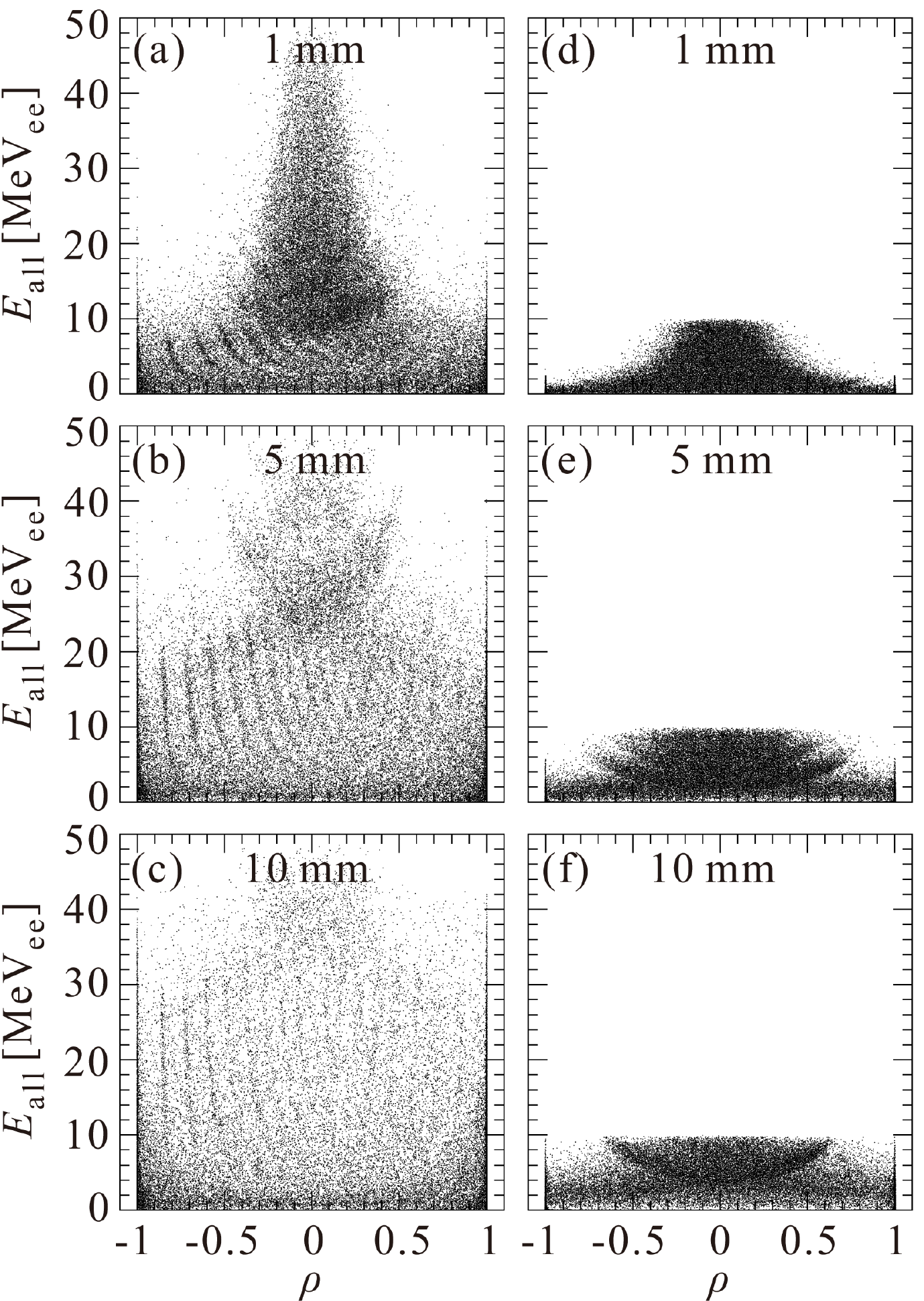}
\caption{\label{SRNG}
The scatter plots for the total pulse height $E_{\rm all}$ in electron-equivalent energy and the balance ratio $\rho$ of all detected events with neutrons of 20 to 60 MeV in (a)--(c) and $\gamma$ rays of 0 to 10 MeV in (d)--(f). The assumed thickness of the layer is shown in each panel.}
\end{figure}

\begin{figure}
\centering
\includegraphics[width=9cm,clip]{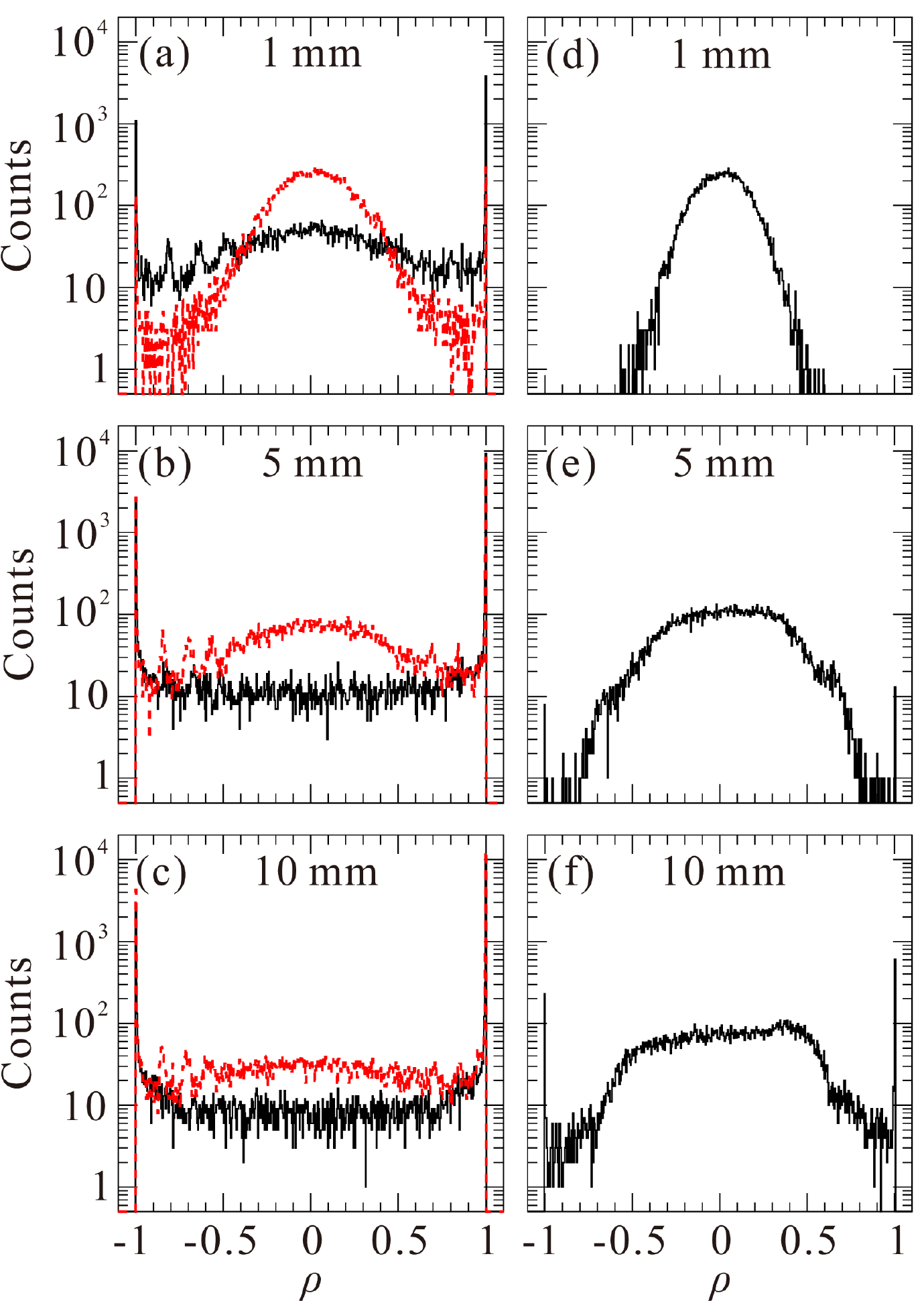}
\caption{\label{RNG}
The $\rho$ distributions for events with $E_{\rm all}$ between 5 and 10 MeV$_{\rm ee}$ (black histograms for neutrons in (a)--(c) and $\gamma$ rays in (d)--(f)) and above 10 MeV$_{\rm ee}$ (red-dashed histograms for neutrons in (a)--(c)), obtained by slicing the corresponding scatter plots in Fig. \ref{SRNG}.}
\end{figure}

To demonstrate the principle of the $\it{n}$-$\gamma$ discrimination, we show the detector responses to neutrons and $\gamma$ rays in Fig. \ref{SRNG}, which are the scatter plots for the total pulse height from the scintillators $E_{\rm all}$ in electron-equivalent energy and the balance ratio $\rho$, defined by Eq. \ref{eq1}, of all detected events obtained with different thicknesses. The $\it{n}$-$\gamma$ discrimination efficiency by $\rho$ can be examined by selecting the same region of energy deposit from 5 to 10 MeV$_{\rm ee}$, as shown by the black histograms in Fig. \ref{RNG}. Clear differences in the $\rho$ distributions for neutrons and $\gamma$ rays are observed. The neutrons are more likely to distribute around $\rho$=$-1$ or 1, whereas most of the $\gamma$ rays distribute in the region around $\rho$=0. These trends demonstrate the practical use of $\rho$ to discriminate neutrons and $\gamma$ rays.

The simplest method of $\it{n}$-$\gamma$ discrimination is to define the neutron and $\gamma$-ray events as follows: an event is identified as a neutron candidate if $\mid$$\rho$$\mid$$>$0.9, otherwise it is regarded as a $\gamma$-ray candidate. This discrimination method, however, has certain probabilities of mis-identification. Depending on the reaction points and layer thicknesses, some neutrons may penetrate more than one layer, resulting in their $\rho$ values being distributed between $-1$ and 1, and thus are mis-identified as $\gamma$ rays. Some low-energy $\gamma$ rays, on the other hand, may appear at around $\rho$=$-1$ or 1, and are incorrectly identified as neutrons. The mis-identifications of neutron and $\gamma$ ray have opposite thickness dependencies. Although the simulations suggest that a lower-threshold operation and a better identification of $\gamma$ rays can be achieved with thinner scintillators, the mis-identification of neutrons increases at the same time. On the contrary, the mis-identification of $\gamma$ rays increases when the layers become too thick for the secondary electron to enter the next layer, although thicker layers do help to provide better identification of neutrons. Furthermore, the identification efficiencies for neutron and $\gamma$ ray depend strongly on the energy deposit, as can be seen in Fig. \ref{SRNG} and the red-dashed histograms in Fig. \ref{RNG} (a)--(c). Namely, better $\gamma$-ray but worse neutron identifications are observed for all thicknesses, with increased energy deposit.

The appropriate layer thickness should not only offer a clear separation between neutrons and $\gamma$ rays, but also provide a good compromise between mis-identifications of $\gamma$ rays and neutrons. On one hand, good $\gamma$-ray identification (i.e. suppression) at a low threshold is important to optimize the purity of the identified neutrons. On the other hand, mis-identification of neutrons should be minimized (i.e. neutron survival rate should be maximized) to ensure sufficient efficiency. Thus the thickness should be optimized to achieve minimal mis-identifications of neutrons and $\gamma$ rays at thresholds as low as reasonably possible.

\subsubsection{Efficiency of $n$-$\gamma$ discrimination and determination of layer thickness}
\label{2.2.2}
\begin{figure}
\centering
\includegraphics[width=11.5cm,clip]{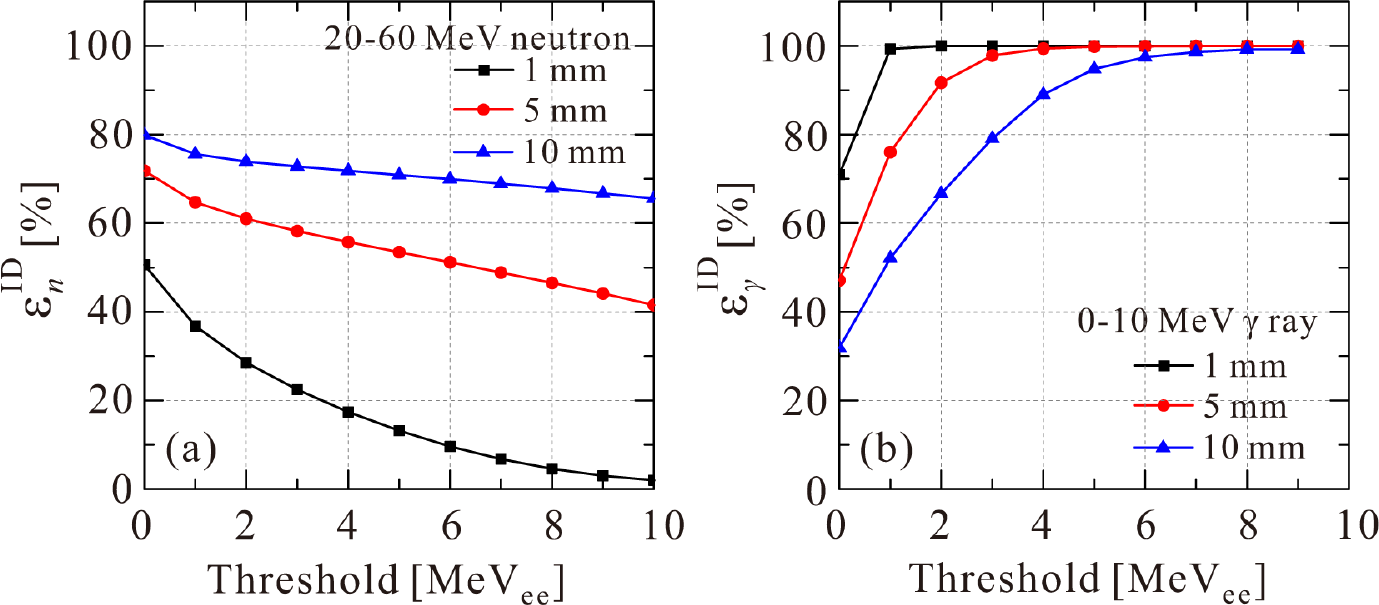}
\caption{\label{1D}
The identification efficiencies of (a) neutrons $\epsilon_{n}^{\rm ID}$ and (b) $\gamma$ rays $\epsilon_{\gamma}^{\rm ID}$ as functions of pulse height threshold with different thicknesses of layers. See text for details.}
\end{figure}

To evaluate the performances for different layer thicknesses quantitatively, we calculated the efficiencies of $n$-$\gamma$ discrimination. The detected particles in the simulations were identified using the $\rho$ difference defined above at different pulse height thresholds. The identification efficiencies of neutrons $\epsilon_{n}^{\rm ID}$ and $\gamma$ rays $\epsilon_{\gamma}^{\rm ID}$ can be expressed as
\begin{eqnarray}
\label{eq2}
\begin{aligned}
\epsilon_{\it n}^{\rm ID} &=N^{\it n}_{\it n}/(N^{\it n}_{\it n}+N^{\it n}_{\gamma}),\\
\epsilon_{\gamma}^{\rm ID} &=N^{\gamma}_{\gamma}/(N^{\gamma}_{\gamma}+N^{\gamma}_{\it n}),
\end{aligned}
\end{eqnarray}
where $N_{b}^{a}$ is the number of detected particle ``$\it a$'' that is identified as ``$\it b$'' at each threshold. Figure \ref{1D} shows the dependence of $\epsilon_{n}^{\rm ID}$ and $\epsilon_{\gamma}^{\rm ID}$ on the pulse height threshold with different thicknesses. It can be inferred that 1-mm thickness is not suitable for our purpose. Although the desired suppression of $\gamma$ rays is achieved at a relatively low threshold ($\epsilon_{\gamma}^{\rm ID}$$>$99$\%$ at 1-MeV$_{\rm ee}$ threshold), most of the neutrons are mis-identified as $\gamma$ rays, resulting in a very poor $\epsilon_{n}^{\rm ID}$. The 10-mm thickness design seems to offer a better option with a rather high $\epsilon_{n}^{\rm ID}$ at all thresholds, but the $\it{n}$-$\gamma$ discrimination is not satisfactory at low thresholds, namely $\epsilon_{\gamma}^{\rm ID}\le$90 $\%$ below 5-MeV$_{\rm ee}$ threshold. The 5-mm thickness design offers a balanced solution with 99.9$\%$ $\gamma$-ray suppression, and 53.5$\%$ neutron survival at 5-MeV$_{\rm ee}$ threshold.

The neutron survival efficiency is relatively low using the above discrimination method. Such situation occurs because the separation by $\rho$ strongly depends on the region of energy deposit, and mis-identification of neutrons is biased at high energy deposit, as mentioned in Sec. \ref{2.2.1}. For practical use, it is necessary to determine a two-dimensional discrimination cut in a $E_{\rm all}$-$\rho$ plot. Particle selection can be made by a discrimination curve $f(\rho)$. An event with $E_{\rm all}$$<$$f(\rho)$ is identified as a $\gamma$-ray candidate, whereas the one with $E_{\rm all}$$\ge$$f(\rho)$ is identified as a neutron candidate.

\begin{figure}
\centering
\includegraphics[width=11.5cm,clip]{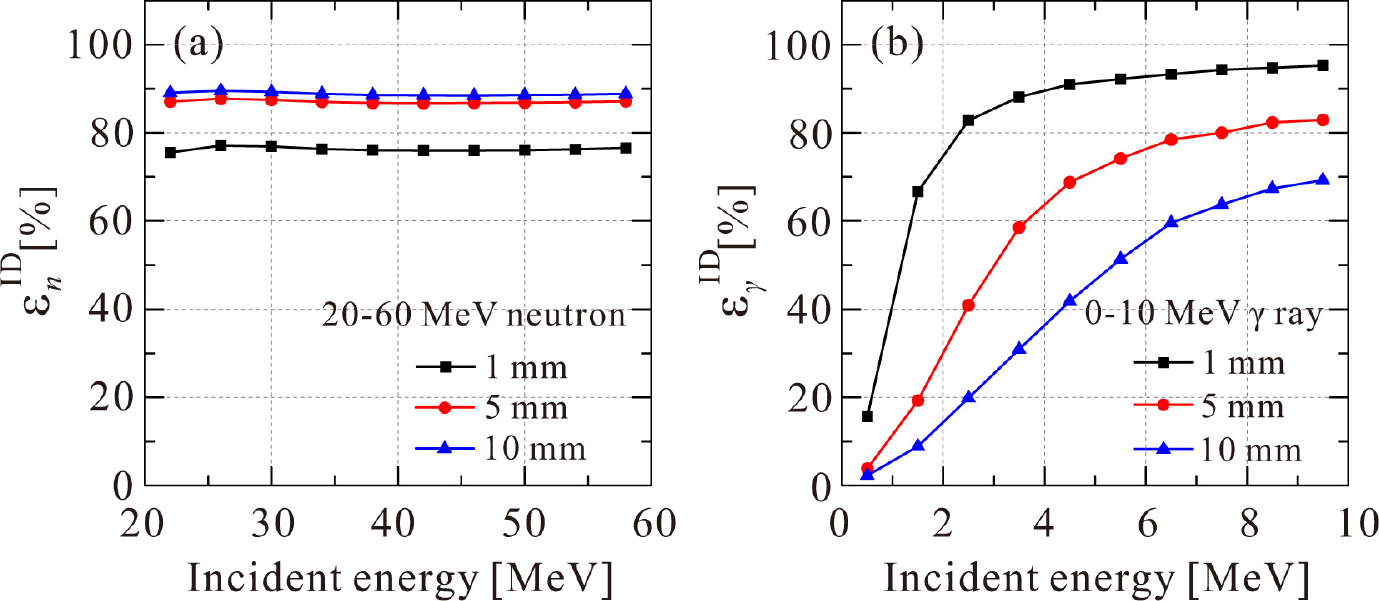}
\caption{\label{2D}
The identification efficiencies of (a) neutrons $\epsilon_{n}^{\rm ID}$ and (b) $\gamma$ rays $\epsilon_{\gamma}^{\rm ID}$ as functions of incident energy with a rectangular discrimination cut of $\rho^{\rm cut}$=0.9 and $E_{\rm all}^{\rm cut}$=10 MeV$_{\rm ee}$. No pulse height threshold is applied here. See text for details.}
\end{figure}

For simplicity, we adopted a rectangular cut defined by $\rho$=$\pm$$\rho^{\rm cut}$ and $E_{\rm all}$=$E_{\rm all}^{\rm cut}$ for the two-dimensional discrimination. We investigated the $\epsilon_{\gamma}^{\rm ID}$ and $\epsilon_{n}^{\rm ID}$ as functions of incident energy with $\rho^{\rm cut}$=0.9 and $E_{\rm all}^{\rm cut}$=10 MeV$_{\rm ee}$, as shown in Fig. \ref{2D}. The value in the $x$-axis is the mean value per incident-energy step (4-MeV step for neutrons and 1-MeV step for $\gamma$ rays, respectively). No pulse height threshold was applied in this analysis. One sees immediately improved $\epsilon^{\rm ID}_n$'s for all thicknesses. $\epsilon^{\rm ID}_n$ improves considerably with thickness from about 76$\%$ at 1 mm to about 87$\%$ at 5 mm, but increases only marginally after that. Note also that for neutrons with incident energies from 20 to 60 MeV, $\epsilon^{\rm ID}_n$'s do not have strong dependence on the incident energy because the proportion of the mis-identified neutrons (enclosed by the rectangular cut), which mainly result from the conversions near the interfaces between layers, depends less on the incident energy, but more on the geometry of the detector. The $\epsilon^{\rm ID}_\gamma$ values, on the other hand, decrease with thickness. Similar to the case with one-dimensional ($\rho$) discrimination, thinner layers are preferable to achieve better $\gamma$-ray suppression and operation at a lower threshold.

Therefore, considering the optimal $\epsilon^{\rm ID}_n$ and $\epsilon^{\rm ID}_\gamma$, we have chosen to use 5-mm-thick plastic scintillators for our prototype S$^{4}$ detector. We note that the selection of $E^{\rm cut}_{\rm all}$ = 10 MeV$_{\rm ee}$, which is the maximal energy deposit attributed to de-excitation $\gamma$ rays, is sufficient in the above example for the following reason. Below 10 MeV$_{\rm ee}$, most of the neutrons concentrate at around $\rho$ = $\pm 1$, and those within $| \rho |$ $<$ 0.9 are distributed uniformly for all layer thicknesses (see Fig. 3). Hence, a further reduction in $E^{\rm cut}_{\rm all}$ will not alter the conclusion. In practical use, however, the values of $E_{\rm all}^{\rm cut}$ and $\rho^{\rm cut}$ as well as the shape of the two-dimensional discrimination cut need to be optimized based on the realistic experimental conditions, i.e. energy distributions and fluxes of neutrons and $\gamma$ rays, as well as the detector response.

From Fig. \ref{2D} (b), we have realized that it is difficult to separate $\gamma$ rays of very low energies from neutrons, which is a limitation of this technique, because a low-energy secondary electron behaves just like a proton, i.e. the electron is stopped in the same layer where it is generated. We note that further suppression of $\gamma$ rays, especially the prompt $\gamma$ rays, may be achieved by considering the TOF measurement. The details of $n$-$\gamma$ discrimination with the experimental data will be described in Sec. \ref{4.3}.

\section{Detector configuration and readout system}
\label{3}
\begin{figure}
\centering
\includegraphics[width=3.5cm,clip]{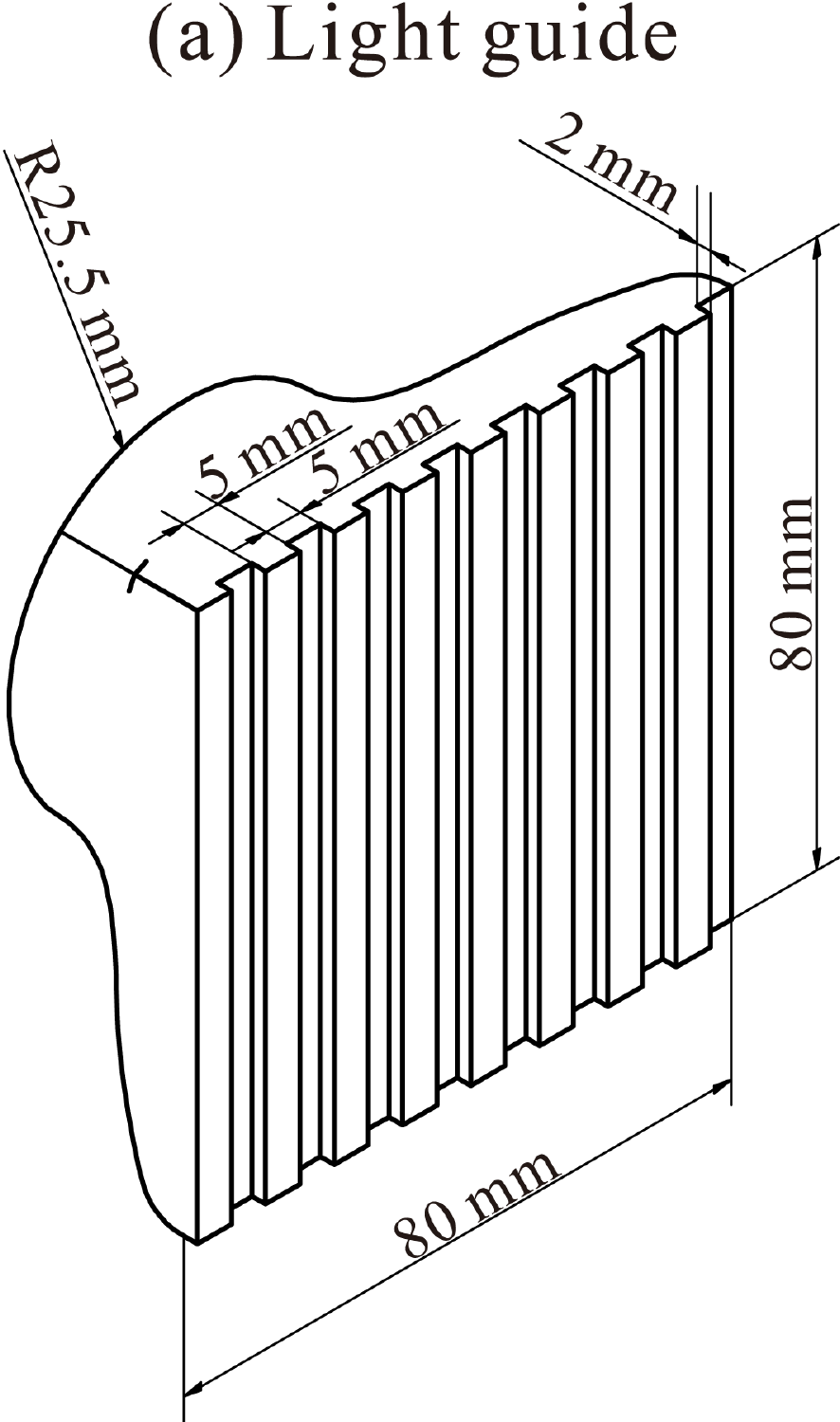}
\includegraphics[width=7.3cm,clip]{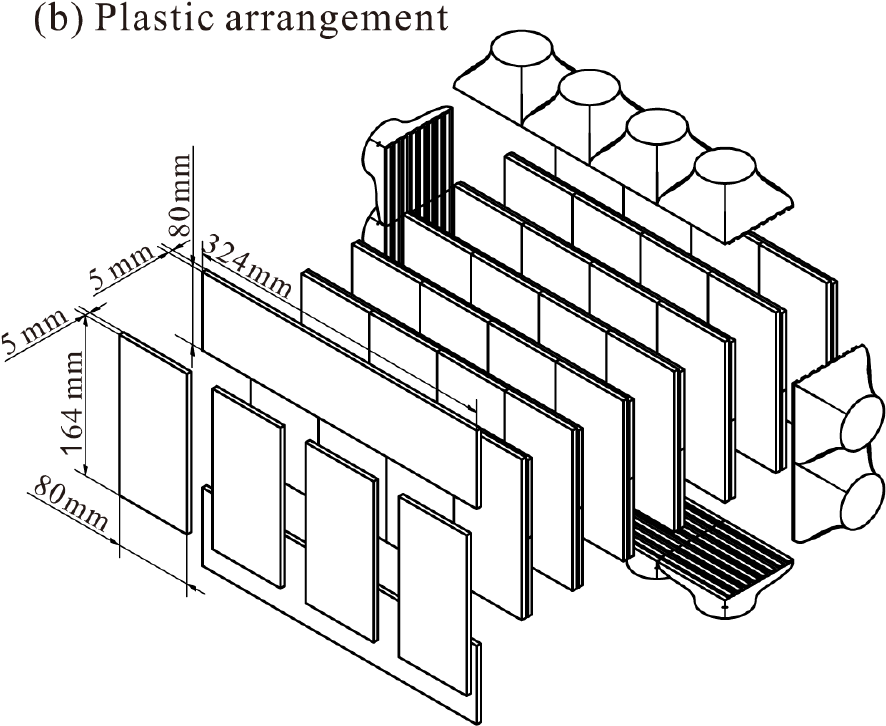}
\includegraphics[width=6.5cm,clip]{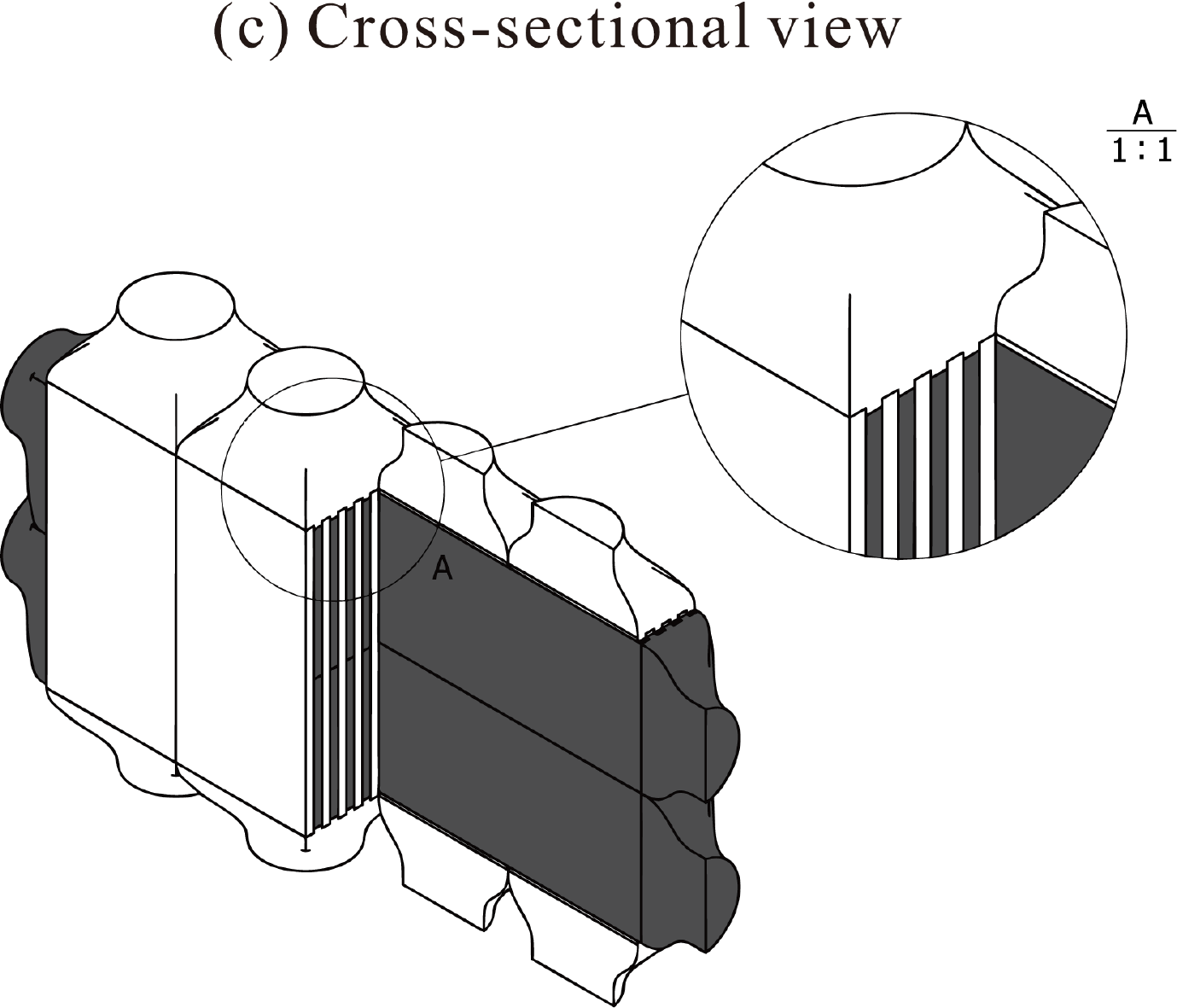}
\caption{\label{detector}
Assembly diagrams of (a) the light guide and (b) plastic arrangement. (c) Cross-sectional view of the assembled BOS$^4$ detector.
In the cross-sectional view, the odd and even layers as well as the attached light guides are displayed in white and grey, respectively.}
\end{figure}

The constructed neutron detector, named Beihang-Osaka university Stack Structure Solid organic Scintillator (BOS$^4$) detector, consists of two sizes of 5-mm-thick BC408-equivalent plastic scintillator plates, stacked together to form a sixteen-layer scintillator with a total thickness of 80 mm. Figure \ref{detector} shows the schematic view of the BOS$^4$ detector. The dimensions of the detector are 320 mm and 160 mm in horizontal and vertical directions, respectively. An odd layer is horizontally segmented into four plastic scintillator plates, each with an active area of 80$\times$160 mm$^{2}$. An even layer is vertically segmented into two plates, each with a 320$\times$80 mm$^{2}$ active area. Each plate is wrapped with a 12-$\mu$m-thick aluminized Mylar along the long sides to reflect light and prevent light leakage into adjacent plate(s). To read out the odd or even layers collectively, the short (80 mm) ends of the odd or even plates are attached to light guides by optical grease, as shown in Fig. \ref{detector}.

The segmentation of odd and even layers of the BOS$^4$ detector allows position determination; the detector has eight sub-sections defined by four sets of odd (vertical) plates denoted as 1o, 2o, 3o, 4o and two sets of even (horizontal) plates as 1e, 2e. Every sub-section can be used as a self-contained unit for neutron detection and $\it{n}$-$\gamma$ discrimination. Moreover, such segmentation offers a good means to reduce background by utilizing the ``firing'' information in the odd and even layers. Detailed discussions are given in Sec. \ref{4.1} and \ref{4.2}. In the following text, 1o-4o and 1e-2e are defined as the odd and even sub-components, respectively.

All sub-components are read out by six pairs of photomultipliers (PMTs; Hamamatsu H7195/H6410); the odd sub-components are read out by four pairs of PMTs connected vertically, and the even sub-components by two pairs of horizontal PMTs attached at both ends of the long side. The analogue output of every PMT was divided into two parts, one of which was fed into a Fast Encoding and Readout Analogue-to-Digital Converter (FERA; LeCroy 4300B) module and the other one into a constant fraction discriminator (CFD; ORTEC 935) to generate timing signals. One of the CFD outputs was sent to a Time-to-Digital Converter (TDC) system which consisted of Time-to-FERA Converter (TFC; LeCroy 4303) and FERA modules. CFD output signals from both ends of the same sub-component were fed into a Mean Timer (REPIC; RPN-070) module to generate a coincident signal. The timings of the two signals were averaged to minimize the position dependence of the output coincidence timing. The trigger signal of the BOS$^{4}$ detector was made by the logic OR of the coincident signals of the six sub-components.

\section{Experiment}
\label{4}
The BOS$^4$ detector was tested using cosmic rays and ion beams in a series of experiments at the cyclotron facility of Research Center for Nuclear Physics (RCNP), Osaka University. Cosmic rays and protons produced by the $^{12}$C($\it{p}$,$\it{pd}$) reaction were used to calibrate the light output. Monoenergetic neutrons generated by $\it{d}$+$\it{d}\to\it{n}$+$^{3}$He reaction, denoted simply as ($d$,$^3$He) hereafter, were used to measure the neutron detection efficiency as well as to investigate the $\it{n}$-$\gamma$ discrimination. It is worth noting that the $^{12}$C($\it{p}$,$\it{pd}$) and ($d$,$^3$He) experiments were performed at the recently constructed Grand Raiden Forward mode beam line (GRAF) \cite{GRAF}. The use of the GRAF beam line, of which the Faraday cup is located more than 20 meters downstream from the scattering chamber, has helped to reduce the $\gamma$-ray background significantly.

\subsection{Experimental setup}
\label{4.0}
\begin{figure}
\centering
\includegraphics[width=9cm,clip]{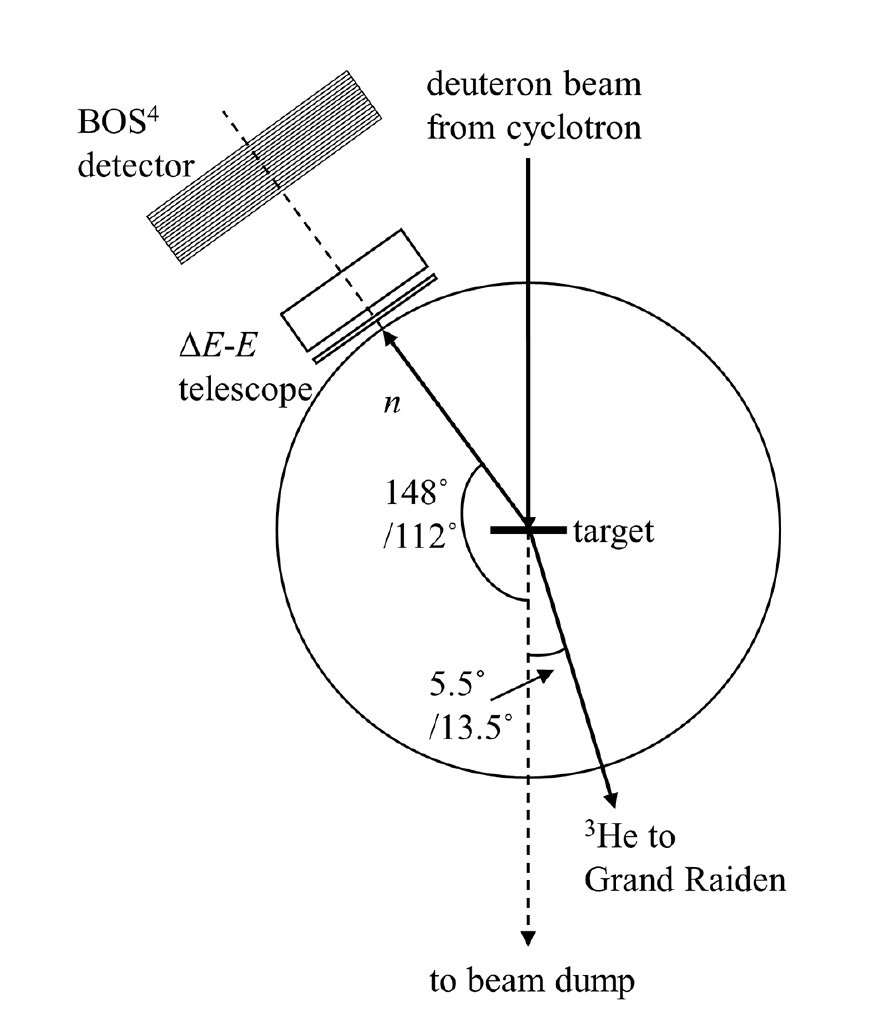}
\caption{\label{setup}
Schematic view of the experimental setup around the target for the ($d$,$^3$He) measurements at 5.5$^{\circ}$ and 13.5$^{\circ}$.}
\end{figure}

In the ($d$,$^3$He) experiment, a deuterated polyethylene (CD$_{2}$) target with a thickness of 24.3 mg/cm$^{2}$ was irradiated by a 196-MeV deuteron beam with an intensity of 10 nA. To obtain monoenergetic neutrons, the scattered $^{3}$He particles at 5.5$^{\circ}$ and 13.5$^{\circ}$ were momentum-analyzed by the Grand Raiden (GR) spectrometer \cite{GR}. Recoil neutrons with kinetic energies centered at 16.5 and 31.2 MeV were detected by the BOS$^4$ detector at the corresponding kinematical central angles of 148$^{\circ}$ and 112$^{\circ}$, respectively. The experimental setup around the target is displayed in Fig. \ref{setup}. A membrane made of 400-${\rm \mu}$m-thick stainless steel, which was installed on the scattering chamber for vacuum isolation, worked as a shield to eliminate low-energy electron background from the target. A set of $\Delta E$ and $E$ plastic scintillation detectors, with thicknesses of 3 and 60 mm respectively, was placed between the CD$_{2}$ target and the BOS$^4$ detector mainly for other experimental purpose which will not be discussed in this article. In the present work, the thin $\Delta E$ scintillator was used as the veto detector to reject charged particles. The total active area of the telescope system is 240$\times$90 mm$^{2}$ with the same angular coverage as the BOS$^4$ from the target. The flight paths from the target to the surface of the $\Delta E$, $E$ and BOS$^4$ detectors for both settings were 45, 49 and 71 cm, respectively. The kinetic energies of the neutrons were determined by the TOF between the target and the BOS$^4$ detector. The details are described later in Sec. \ref{4.2}.

\subsection{Light output calibration}
\label{4.1}
To define the energy threshold so as to determine the neutron detection efficiency, we calibrated the light output in unit MeV$_{\rm ee}$. The light output ($Q^{i{\rm o(e)}}$; $i$=1,2,3,4 (or 1,2)) measured in the $i$-th odd (even) sub-component is constructed by taking the geometric mean of the analogue signals from the PMT outputs at both ends, $Q^{i{\rm U(L)}}$  and $Q^{i{\rm D(R)}}$, i.e.
\begin{equation}
\label{eq3}
Q^{i{\rm o(e)}}=\sqrt{Q^{i{\rm U(L)}}\cdot Q^{i{\rm D(R)}}},
\end{equation}
where U (D) and L (R) represent the upper (lower) signal for the odd and the left (right) signal for the even sub-components, respectively.

The relationship between the deposited energy in MeV$_{\rm ee}$ and the ADC outputs $Q^{i{\rm o(e)}}$ was calibrated for each sub-component. For the BOS$^{4}$ detector, it is necessary to lower the pulse height thresholds for odd and even layers to observe the Compton edge of $\gamma$ rays. Such low threshold condition is hard to fulfill for most $\gamma$ rays. Therefore, a usual calibration method using standard $\gamma$-ray sources is not practical. In the present work, we used cosmic rays and intermediate-energy protons to study the energy response of the detector.

\begin{figure}
\centering
\includegraphics[width=11.5cm,clip]{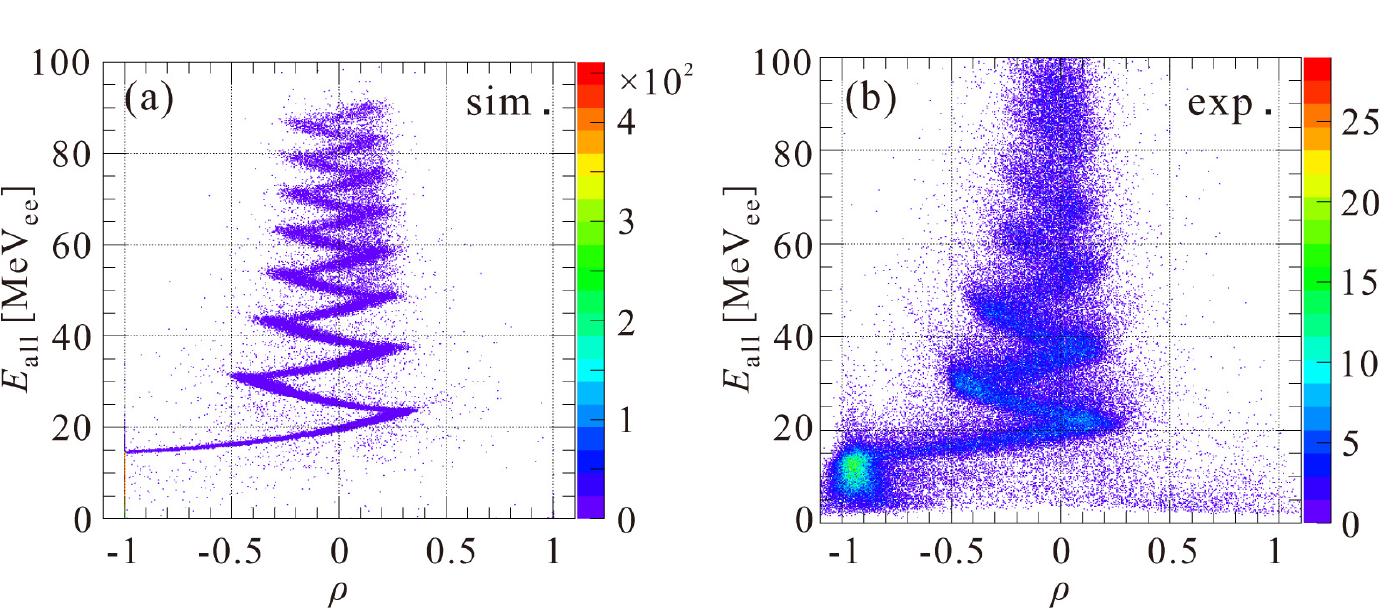}
\caption{\label{proton}
The zig-zag structure of the $E_{\rm all}$-$\rho$ plots for protons with continuous energy in (a) simulation (sim.) and (b) experimental data (exp.) after calibration. The simulation result is the original output without any adjustment.}
\end{figure}

The cosmic rays were measured with the BOS$^4$ detector laid face up on a table. Using the observed peak of muons, we adjusted the relative gain between two PMTs at both ends of each sub-component. Next, we used the protons with continuous energy up to 140 MeV, produced by the $^{12}$C($\it{p}$,$\it{pd}$) reaction. The $E$ detector in front of the BOS$^4$ was removed during this measurement. When charged particles with continuous energy enter the BOS$^4$ detector, the deposited energies are distributed among all of the penetrated layers, resulting in a zig-zag structure in the $E_{\rm all}$-$\rho$ plot. Figure \ref{proton}(a) shows the $E_{\rm all}$-$\rho$ plot obtained with a Monte Carlo simulation using the Geant4 code taking into account the realistic geometry of the experimental setup. Each turning point in the plot indicates that the proton of a certain energy stops at the end of a certain layer. Here, the theoretical light output of the maximum stopping energy in each layer was estimated by identifying each turning point. The zig-zag structure in the experimental data was observed for each sub-section by using $Q^{i{\rm o}}$ and $Q^{i{\rm e}}$. Since we had no means to calibrate the light output layer by layer, a linear coefficient $C^{i{\rm o(e)}}$ between the theoretical values (MeV$_{\rm ee}$) and ADC outputs (channel) was determined for the $i$-th odd(even) sub-component based on the first eight turning points. The turning points for the deeper layers are hard to identify in the experimental data due mainly to the energy resolution and multiple scattering effect.

After the calibration of each sub-component, the sum of the energy deposits from the constitutive odd (even) sub-components is taken as the total energy deposit in the odd (even) layers $E_{\rm odd (even)}$. In determining $E_{\rm odd(even)}$, we have suppressed background signals by taking advantage of the layer segmentation as described below. We consider the $i$-th odd (even) sub-component to be ``fired'', denoted as $F^{i{\rm o(e)}}$=1, if the recorded timings from both ends are within the reasonable range; otherwise it is considered as ``unfired'' and denoted as $F^{i{\rm o(e)}}$=0. At a normal event rate, only one of the odd and/or even sub-components (one of the sub-sections) can be fired except for some events that occur near the borders of adjacent sub-components. Instead of summing up all sub-components and in the process adding electronic noise from the unfired ones, we determine the $E_{\rm odd(even)}$ by adding the outputs of the constitutive odd (even) sub-components depending on their firing conditions, namely,
\begin{eqnarray}
\label{eq4}
\begin{aligned}
E_{\rm odd}&=\sum_{i=1}^{4}C^{i{\rm o}} Q^{i{\rm o}} F^{i{\rm o}},\\
E_{\rm even}&=\sum_{i=1}^{2}C^{i{\rm e}} Q^{i{\rm e}} F^{i{\rm e}}.
\end{aligned}
\end{eqnarray}
In the case where only the odd (even) sub-components are fired ($E_{\rm even(odd)}$=0), the light output of the even (odd) sub-component with the highest pulse height is taken as $E_{\rm even(odd)}$. The definitions of $E_{\rm all}$ and $\rho$ are given by Eq. \ref{eq1}. Unless otherwise stated, the $E_{\rm all}$-$\rho$ plots shown throughout this article refer to the sum of all fired sub-component(s).

The zig-zag structure of the $E_{\rm all}$-$\rho$ plot for the measurement of protons is shown in Fig. \ref{proton}(b). Minor differences are observed for some turning points that may be attributed to the difference of light collection between layers, due probably to the contact condition with the light guides. Such differences are considered and discussed when determining the energy resolution in the following section.

\subsection{Energy, timing and position resolutions}
\label{4.2}
\begin{figure}
\centering
\includegraphics[width=11.5cm,clip]{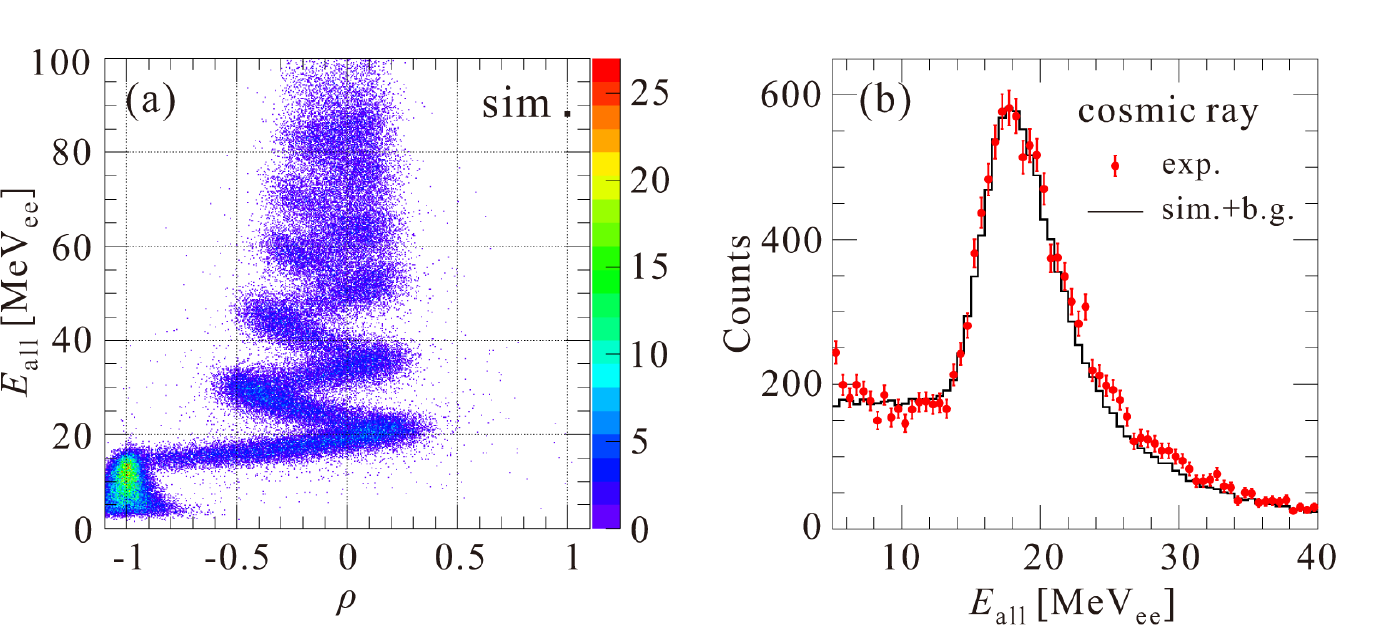}
\caption{\label{resolution}
(a) The simulated (sim.) $E_{\rm all}$-$\rho$ plot for protons and (b) the experimental (exp.) and simulated (sim.) light outputs of cosmic rays. The energy resolution and light collection fluctuations of layers are considered in the simulations. An additional exponential background (b.g.) is added to the simulated response function of cosmic rays. See text for details.}
\end{figure}

The intrinsic energy resolution $\sigma(L)$ of a scintillator attributed to photon statistics can be simply described as follows \cite{Leo},
\begin{equation}
\label{eq5}
\frac{\sigma(L)}{L}=\frac{\alpha}{\sqrt{L}},
\end{equation}
where $L$ is the light output in MeV$_{\rm ee}$, and $\alpha$ is a proportionality factor. As mentioned in the last section, there exists light collection difference between layers due possibly to the contact problem. Such fluctuations of detected photon number also contribute to the spread of the total light output $E_{\rm all}$. Therefore, we have considered both contributions in the simulations to fully understand the energy resolution of the experimental data. The proportionality factor $\alpha$ of the intrinsic resolution was determined to be 0.33 by fitting the width of the output from the first and second layers in Fig. \ref{proton}(b). We assume that $\alpha$ is a common factor for all layers. Next, we introduced photon number fluctuations for layers to the original outputs of the simulation. The fluctuation of each of the shallower eight layers was determined by comparing the simulated and experimental turning points in Fig. \ref{proton}(a) and (b), respectively. The fluctuations of the deeper layers are assumed to be the same as the shallower ones. After the adjustment in the simulation, the thickness of the zig-zag line is well reproduced, as shown in Fig. \ref{resolution}(a). The total light output response of all layers is simulated for the cosmic rays assuming only muons, and compared with the experimental data, as displayed in Fig. \ref{resolution}(b). Here, we have added an exponential background in addition to the simulated response function of cosmic rays. The peak position and structure of cosmic rays with a Landau tail are well reproduced, which confirms the validity of the energy calibration.

The time steps of the TDC outputs were calibrated by the radio-frequency (RF) period of the cyclotron in the ($d$,$^3$He) experiment. The timing signals from two PMTs attached to both ends of the $i$-th odd(even) sub-component are denoted as $t^{i{\rm U(L)}}$ and $t^{i{\rm D(R)}}$. For simplicity, we omit the index $i$ hereafter. The width of the time difference between $t^{\rm U(L)}$ and $t^{\rm D(R)}$ is about 300 psec in $\sigma$. It was estimated by restricting the spatial distribution of the neutrons (to minimize contribution from the position dependence) using two-body kinematics with the position spectrum of $^{3}$He. The width reflects the intrinsic time resolution of the detector associated with the PMTs and electronics.

The TOF information from the reaction target to each odd(even) sub-component $t^{\rm o(e)}$ is determined by the average of $t^{\rm U(L)}$ and $t^{\rm D(R)}$ as well as the RF signal $t_{\rm RF}$, i.e.
\begin{equation}
\label{eq6}
t^{\rm o(e)}=\frac{t^{\rm U(L)}+t^{\rm D(R)}}{2}-t_{\rm RF}+t_{0}^{\rm o(e)},
\end{equation}
where $t_{0}^{\rm o(e)}$ is an offset, which can be calibrated by the prompt $\gamma$ rays for each sub-component. Slewing corrections using the pulse height information of the prompt $\gamma$ rays were applied for this analysis. The typical TOF resolution of prompt $\gamma$ rays is 350 psec in $\sigma$, which includes the intrinsic time resolution, the time fluctuation of the RF signal and the effect of finite thickness of the detector. All of these components can be consistently understood. The width of TOF corresponds to energy resolution of about 8$\%$ for neutron kinetic energy of 40 MeV.

To deduce the angular distribution of neutrons, information of the hit position on the detector is necessary. The position on each sub-component $x^{\rm o(e)}$ is given by the difference between $t^{\rm U(L)}$ and $t^{\rm D(R)}$, as expressed below,
\begin{equation}
\label{eq7}
x^{\rm o(e)}=c_{\rm eff}\frac{t^{\rm U(L)}-t^{\rm D(R)}}{2}+x_{0}^{\rm o(e)},
\end{equation}
where $c_{\rm eff}$ is the effective propagation speed of light in the detector, and $x_{0}^{\rm o(e)}$ is a calibration constant for each sub-component. $c_{\rm eff}$ is about 16 cm/ns which was deduced from the distribution of time difference $t^{\rm U(L)}-t^{\rm D(R)}$ (about 2 nsec difference in 32-cm distance). The position resolution derived from the time difference width of 300 psec is around 2.4 cm in $\sigma$. Firing of both odd and even sub-components allows a two-dimensional position determination. In a case that only the odd (even) sub-component is fired, the position in the orthogonal direction is determined by the segmentation.
\subsection{Neutron detection efficiency determined by ($\it{d}$,$^{3}$He) measurement}
\label{4.3}
To measure the neutron detection efficiency, we irradiated the BOS$^4$ detector with monoenergetic neutron beams centered at two different energies produced by the $\it{d}$+$\it{d}\to\it{n}$+$^{3}$He reaction. The measurements covered two kinetic energy regions from 15.4 to 17.6 MeV and from 28.6 to 33.8 MeV at 5.5$^\circ$ and 13.5$^\circ$ settings of the ($\it{d}$,$^{3}$He) experiment, respectively. The detection efficiency was determined by the ratio of the number of detected and identified neutrons in coincidence with $^3$He to the number of the incoming neutron flux, defined by the $^{3}$He events. The $^{3}$He particles were detected and identified using the GR spectrometer. By employing $\rho$ and TOF information simultaneously, mis-identified $\gamma$ rays were estimated and subtracted, and the number of neutrons in coincidence with the $^3$He events was deduced. In the following paragraphs, we describe two deduction methods to check the consistency and to estimate the systematic error of neutron events.

\begin{figure}
\centering
\includegraphics[width=11cm,clip]{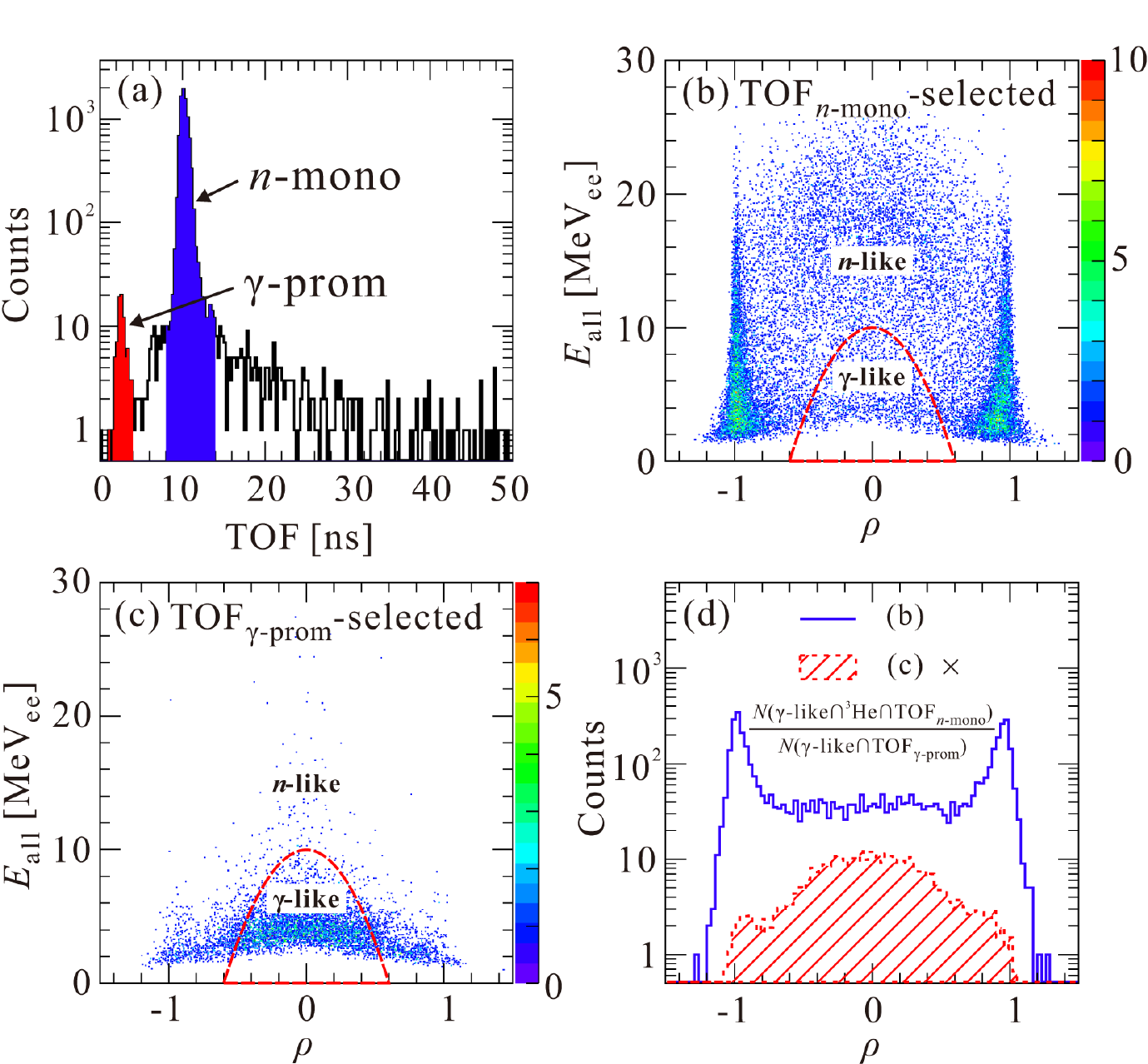}
\caption{\label{discrimination}
(a) Typical TOF spectrum with $^{3}$He coincidence at 13.5$^{\circ}$ setting in the ($\it{d}$,$^{3}$He) experiment. Monoenergetic neutrons ($n$-mono) and prompt $\gamma$ rays ($\gamma$-prom) are indicated by the blue and red regions, respectively. $E_{\rm all}$-$\rho$ plots of (b) neutrons and (c) $\gamma$ rays, obtained by selecting the corresponding regions, denoted by TOF$_{n{\mhyphen}{\rm mono}}$ and TOF$_{\gamma{\mhyphen}{\rm prom}}$, in the TOF spectra. For practical reason, we show the $^3$He-gated and $^3$He-ungated plots in (b) and (c), respectively. The red-dashed lines in (b) and (c) describe the two-dimensional discrimination cut of neutrons and $\gamma$ rays. The ``$n$-like'' region represents all events outside the ``$\gamma$-like'' region. (d) The $\rho$ distributions of neutrons (blue histogram) and $\gamma$ rays (red-dashed histogram with hatched area), obtained by the projections of the plot in (b) and the normalized plot in (c), respectively. See text for details.}
\end{figure}

In the first method, we identified the neutrons using the TOF and estimated the number of mis-identified $\gamma$ rays using $\rho$. Thanks to the low background environment, the monoenergetic neutrons from the $d$($d$,$^3$He) reaction and the prompt $\gamma$ rays from the $^{12}$C($\it{d}$,$^{3}$He) reaction, denoted by ``$n$-mono'' and ``$\gamma$-prom'', respectively, can be easily identified in the $^3$He-gated TOF spectrum for the 13.5$^\circ$ setting, as shown in Fig. \ref{discrimination}(a). For convenience, we define these neutron and $\gamma$-ray events as TOF$_{n{\mhyphen}{\rm mono}}$ and TOF$_{\gamma{\mhyphen}{\rm prom}}$ events, respectively. The $^3$He-gated TOF$_{n{\mhyphen}{\rm mono}}$- and $^3$He-ungated TOF$_{\gamma{\mhyphen}{\rm prom}}$-selected $E_{\rm all}$-$\rho$ plots are shown in Fig. \ref{discrimination}(b) and (c), respectively. The reason for showing the $^3$He-ungated plot in Fig. \ref{discrimination}(c) will become clear later. As expected, the difference in $\rho$ distribution between neutrons and $\gamma$ rays is clearly observed.  The broadenings at $\rho$ around $-1$ and 1 are due to pedestal fluctuation. We should note the possible presence of $\gamma$ rays, which are expected to distribute especially around $\rho$ = 0 at $E_{\rm all}$ below 10 MeV$_{\rm ee}$, in Fig. \ref{discrimination}(b). In this analysis, we defined a two-dimensional discrimination cut for the neutrons and $\gamma$ rays, as shown by the red-dashed parabolic curves in Fig. \ref{discrimination}(b) and (c), with roots $\rho^{\rm cut}$ = $\pm 0.6$ and a vertex point at ($\rho$ = 0, $E^{\rm cut}_{\rm all}$ = 10 MeV$_{\rm ee}$). The events enclosed by the red-dashed lines, denoted by the ``$\gamma$-like'' region, are taken as the $\gamma$-ray-like events, whereas those outside the ``$\gamma$-like'' region as the neutron-like events, denoted by ``$n$-like''. We chose the values of $\rho^{\rm cut}$ and $E_{\rm all}^{\rm cut}$ to cover the $\gamma$ rays in the middle as much as possible, while simultaneously reducing the loss of neutrons at the side regions. The non-hatched histogram in Fig. \ref{discrimination}(d) is the $\rho$ spectrum of the TOF$_{n{\mhyphen}{\rm mono}}$-selected events, which was obtained by the projection of Fig. \ref{discrimination}(b). Assuming that the $\gamma$-like region consists of predominantly $\gamma$-ray events, the number of neutrons $N^n$ was determined by integrating the $\rho$ spectrum of the TOF$_{n{\mhyphen}{\rm mono}}$-selected events after subtracting the accidental $\gamma$ rays as follows:
\begin{equation}
\label{eq8}
\begin{split}
N^n=&\int_{\rho}[N_{\rho}(^{3}{\rm He}\hspace{-0.7 mm}\cap\hspace{-0.7 mm}{\rm TOF}_{n{\mhyphen}{\rm mono}})\\
&-N_{\rho}(^{3}{\rm He}\hspace{-0.7 mm}\cap\hspace{-0.7 mm}{\rm TOF}_{\gamma{\mhyphen}{\rm prom}})\times
\frac{N(\gamma{\mhyphen}{\rm like}\hspace{-0.7 mm}\cap^{3}\hspace{-0.7 mm}{\rm He}\hspace{-0.7 mm}\cap\hspace{-0.7 mm}{\rm TOF}_{n{\mhyphen}{\rm mono}})}{N(\gamma{\mhyphen}{\rm like}\hspace{-0.7 mm}\cap^{3}\hspace{-0.7 mm}{\rm He}\hspace{-0.7 mm}\cap\hspace{-0.7 mm}{\rm TOF}_{\gamma{\mhyphen}{\rm prom}})}]d\rho,
\end{split}
\end{equation}
where the terms $N$ with and without the subscript ``$\rho$'' represent the $\rho$ distributions and $\rho$-integrated events obtained with the conditions in the brackets, respectively. The accidental $\gamma$ rays are given by the second integral on the right hand side of Eq. \ref{eq8}. Since the $E_{\rm all}$-$\rho$ distributions for all $^3$He-gated or $^3$He-ungated $\gamma$ rays, including the prompt $\gamma$ rays, are expected to be almost similar, i.e. $\frac{N(\gamma{\mhyphen}{\rm like}\cap^{3}{\rm He}\cap{\rm TOF}_{\gamma{\mhyphen}{\rm prom}})}{N(\gamma{\mhyphen}{\rm like}\cap{\rm TOF}_{\gamma{\mhyphen}{\rm prom}})}$$\cong$$\frac{N_{\rho}(^{3}{\rm He}\cap{\rm TOF}_{\gamma{\mhyphen}{\rm prom}})}{N_{\rho}({\rm TOF}_{\gamma{\mhyphen}{\rm prom}})}$, we have replaced $\frac{N_{\rho}(^{3}{\rm He}\cap{\rm TOF}_{\gamma{\mhyphen}{\rm prom}})}{N(\gamma{\mhyphen}{\rm like}\cap^{3}{\rm He}\cap{\rm TOF}_{\gamma{\mhyphen}{\rm prom}})}$ in Eq. \ref{eq8} by $\frac{N_{\rho}({\rm TOF}_{\gamma{\mhyphen}{\rm prom}})}{N(\gamma{\mhyphen}{\rm like}\cap{\rm TOF}_{\gamma{\mhyphen}{\rm prom}})}$ to reduce statistical uncertainty. The hatched histogram in Fig. \ref{discrimination}(d) shows the $\rho$ spectrum of the estimated accidental $\gamma$ rays, which was obtained by multiplying the projected spectrum of Fig. \ref{discrimination}(c) by the normalization factor $\frac{N(\gamma{\mhyphen}{\rm like}\cap^{3}{\rm He}\cap{\rm TOF}_{n{\mhyphen}{\rm mono}})}{N(\gamma{\mhyphen}{\rm like}\cap{\rm TOF}_{\gamma{\mhyphen}{\rm prom}})}$.

\begin{figure}
\centering
\includegraphics[width=11.5cm,clip]{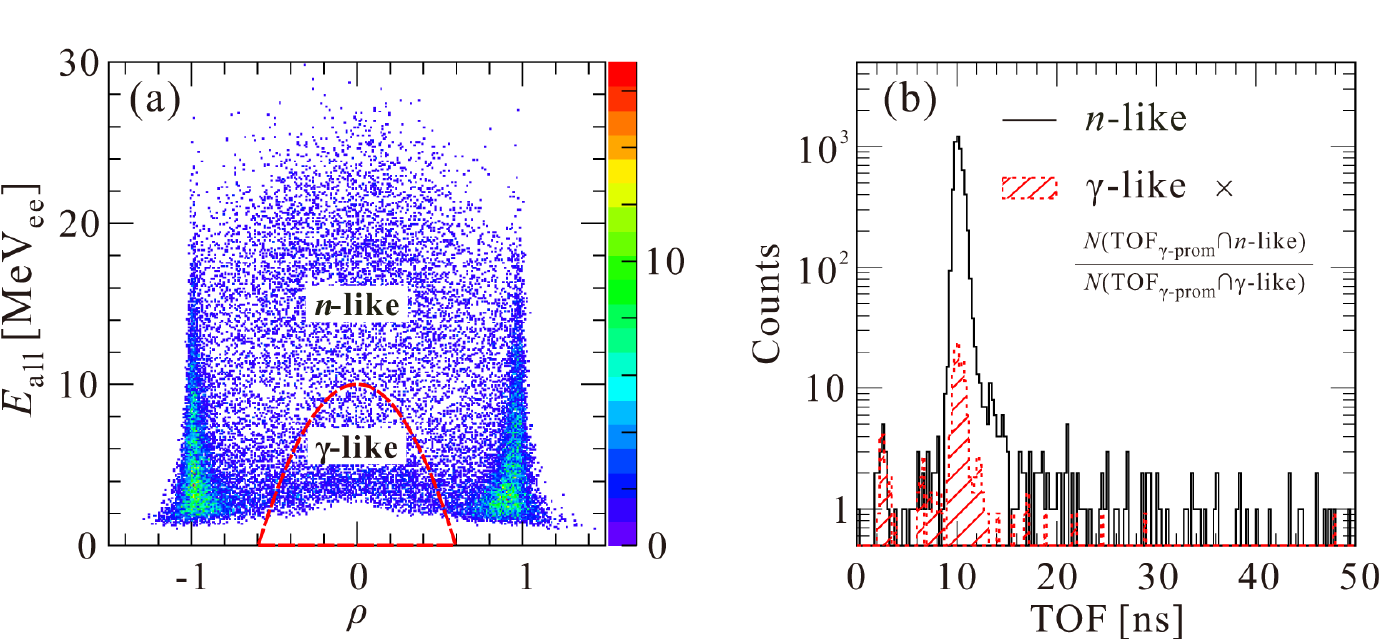}
\caption{\label{TOFsub}
(a) $E_{\rm all}$-$\rho$ plot of the detected neutral-particle events in coincidence with $^3$He for the 13.5$^{\circ}$ setting, with the same two-dimensional discrimination cut as Fig. \ref{discrimination}, shown by the red-dashed line. (b) The corresponding TOF spectra of a sub-component by selecting the two regions. See text for details.}
\end{figure}

The number of neutrons $N^n$ can also be determined using procedures opposite to that in the first method, namely neutrons are selected using the $E_{\rm all}$-$\rho$ information and integrated after subtracting the mis-identified $\gamma$ rays estimated from the TOF spectrum. Figure \ref{TOFsub} (a) shows the $E_{\rm all}$-$\rho$ plot of the detected neutral-particle events in coincidence with $^3$He for the 13.5$^{\circ}$ setting of the ($\it{d}$,$^{3}$He) reaction. Here the same two-dimensional discrimination cut in the first method, shown by the red-dashed parabolic curve, was used. As discussed in Sec. \ref{2.2.2}, the $\it n$-$\gamma$ discrimination by $\rho$ has a non-zero probability of mis-identification of $\gamma$ rays below 5 MeV$_{\rm ee}$, due to the possible mixture of neutrons and $\gamma$ rays. The mis-identified $\gamma$ rays in the $n$-like region of the $E_{\rm all}$-$\rho$ plot can be estimated and subtracted by comparing the TOF spectra of the $n$-like and $\gamma$-like regions as follows:
\begin{equation}
\label{eq9}
\begin{split}
N^n=&\sum_{{\rm sub}{\mhyphen}{\rm component}}\int_{\rm TOF}[N_{\rm TOF}(^{3}{\rm He}\hspace{-0.7 mm}\cap\hspace{-0.7 mm}{n{\mhyphen}{\rm like}})\\
&-N_{\rm TOF}(^{3}{\rm He}\hspace{-0.7 mm}\cap\hspace{-0.7 mm}{\gamma{\mhyphen}{\rm like}})\times
\frac{N({\rm TOF}_{\gamma{\mhyphen}{\rm prom}}\hspace{-0.7 mm}\cap^{3}\hspace{-0.7 mm}{\rm He}\hspace{-0.7 mm}\cap\hspace{-0.7 mm}{n}{\mhyphen}{\rm like})}{N({\rm TOF}_{\gamma{\mhyphen}{\rm prom}}\hspace{-0.7 mm}\cap^{3}\hspace{-0.7 mm}{\rm He}\hspace{-0.7 mm}\cap\hspace{-0.7 mm}{\gamma}{\mhyphen}{\rm like})}]d({\rm TOF}),
\end{split}
\end{equation}
where the terms $N$ with and without the subscript ``TOF'' represent the TOF distributions and TOF-integrated events obtained with the conditions in the brackets, respectively. Since the TOF$_{\gamma{\mhyphen}{\rm prom}}$ events consist mostly of $\gamma$ rays, and the TOF$_{\gamma{\mhyphen}{\rm prom}}$ events found in the $n$-like region are due to the ``leaked'' $\gamma$ rays, we expect the ratios of the leaked $\gamma$ rays to those in the $\gamma$-like region to be similar for the $^3$He-gated and $^3$He-ungated measurements, i.e. $\frac{N({\rm TOF}_{\gamma{\mhyphen}{\rm prom}}\cap^{3}{\rm He}\cap{n}{\mhyphen}{\rm like})}{N({\rm TOF}_{\gamma{\mhyphen}{\rm prom}}\cap^{3}{\rm He}\cap{\gamma}{\mhyphen}{\rm like})}$$=$$\frac{N({\rm TOF}_{\gamma{\mhyphen}{\rm prom}}\cap{n}{\mhyphen}{\rm like})}{N({\rm TOF}_{\gamma{\mhyphen}{\rm prom}}\cap{\gamma}{\mhyphen}{\rm like})}$. Hence, to reduce the statistical uncertainty attributed to the normalization factor in Eq. \ref{eq9}, we determined and adopted the ratio for the $^3$He-ungated measurements. Figure \ref{TOFsub} (b) shows a typical TOF spectrum for the $n$-like region (black histogram) and a normalized TOF spectrum for the $\gamma$-like region (red-dashed histogram with hatched area) of a sub-component with $^{3}$He coincidence. We subtracted the normalized TOF spectrum of the $\gamma$-like region from the TOF spectrum of the $n$-like region with $^{3}$He coincidence, and then integrated the number of net neutrons. The same procedure was applied to the TOF spectrum of each sub-component. The net neutrons were later summed up taking into consideration the event multiplicities, i.e. the number of sub-components that were fired by one event, to avoid double counting.

We deduced the neutron detection efficiencies for different energy thresholds. The values obtained with the two methods are consistent with each other within the statistical uncertainty. The difference between the two, which is much smaller than the statistical error of the neutron events, is taken as the systematic error. The deduction procedures described above were performed for both angular settings. Table \ref{tab1} shows the experimental efficiencies of the BOS$^{4}$ detector at two energies for different thresholds, with each value followed by the statistical and systematic errors, respectively.

\begin{table}
\centering
\caption{\label{tab1} The measured detection efficiencies at two neutron energies for different thresholds from 5 to 10 MeV$_{\rm ee}$. The statistical and systematic errors of the efficiencies are also listed. For simplicity, we show the superscripts that describe the errors only for one data point.}
\begin{tabular}{ c | c c }
\toprule
T$_{n}$ [MeV] &Threshold [MeV$_{\rm ee}$] &Detection efficiency [$\%$]\\
\hline
\multirow{6}{*}{15.4$\sim$17.6}
 &5.0 &3.82$\pm$0.20$^{\rm (stat)}$$\pm$0.04$^{\rm (syst)}$\\
 &6.0 &3.13$\pm$0.18$\pm$0.03 \\
 &7.0 &2.32$\pm$0.16$\pm$0.02 \\
 &8.0 &1.68$\pm$0.16$\pm$0.02 \\
 &9.0 &1.19$\pm$0.11$\pm$0.02 \\
 &10.0 &0.70$\pm$0.08$\pm$0.01 \\
\hline
\multirow{6}{*}{28.6$\sim$33.8}
 &5.0 &6.71$\pm$0.24$\pm$0.02\\
 &6.0 &5.97$\pm$0.22$\pm$0.02 \\
 &7.0 &5.37$\pm$0.21$\pm$0.01 \\
 &8.0 &4.74$\pm$0.20$\pm$0.01\\
 &9.0 &4.25$\pm$0.19$\pm$0.01\\
 &10.0 &3.86$\pm$0.17$\pm$0.01\\
\bottomrule
\end{tabular}
\end{table}

\begin{figure}
\centering
\includegraphics[width=6cm,clip]{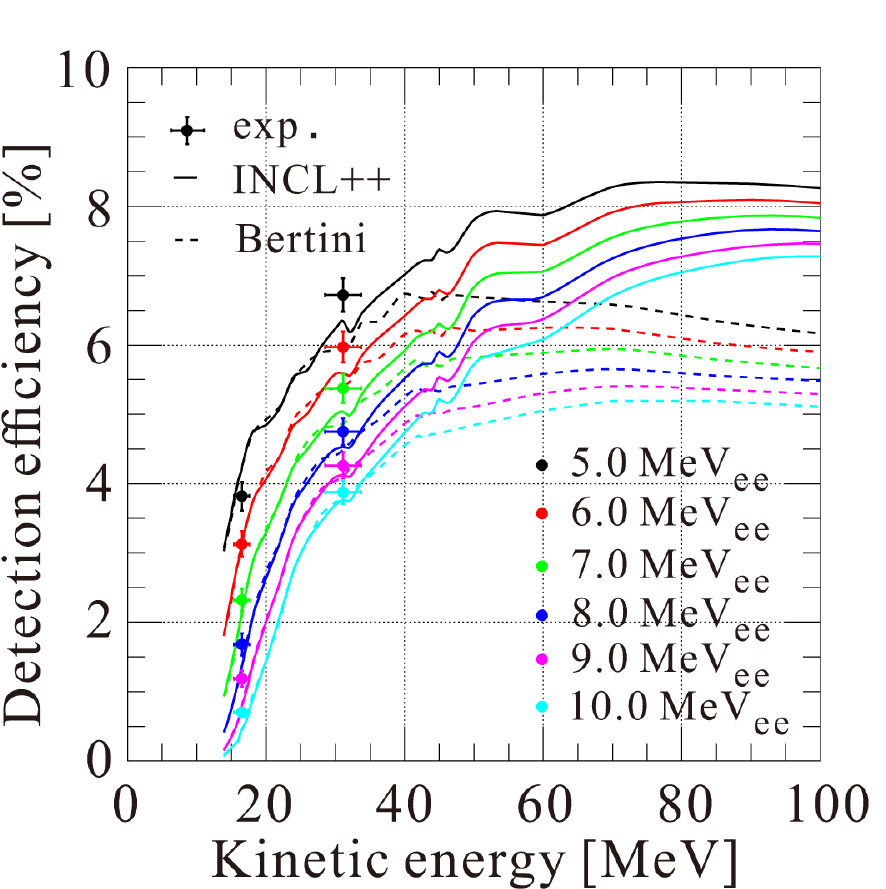}
\caption{\label{efficiency}
Comparison between the measured efficiencies and simulation results for different thresholds.
The solid curves were calculated using the INCL++ model coupled with NeutronHP model,
while the dashed ones were by the Bertini model coupled with NeutronHP model. }
\end{figure}

The experimental efficiencies are compared with the simulations employing the INCL++ model \cite{INCL1,INCL2}. For direct comparison, the same conditions for neutron selection and the subtraction methods were applied in the analysis of the simulation results to include the possible loss of neutrons during the discrimination process. To examine any possible model dependence, we also performed simulations employing the Bertini intranuclear cascade model \cite{BERT}. All materials between the target and the BOS$^4$ detector were included in the simulations. The simulation results using the two models and the experimental data are shown in Fig. \ref{efficiency}. Here, we have taken the quadratic sum of the statistical and systematic errors as the total experimental error. The average kinetic energies at two angular settings are calculated to be 16.5$\pm1.1$ MeV and 31.2$\pm2.6$ MeV, respectively. Good agreements are observed between simulations and experiment at 16.5$\pm1.1$ MeV. At energy region beyond 30 MeV, systematic discrepancies are seen between the two models. Both models agree with the experimental data at 31.2$\pm2.6$ MeV within one to three standard deviations. Due to the small difference between the two models at this energy and the sizable experimental error bars, it is difficult to decide the appropriate model with the present data. Designated measurement of neutron detection efficiency is needed to determine absolute cross sections involving higher-energy neutrons in the future.

\section{Summary}
\label{5}
A multi-layer plastic scintillation detector, the S$^4$ detector, is proposed for the detection of neutrons with kinetic energy up to 100 MeV. We constructed a sixteen-layer prototype detector with a 5-mm layer thickness, named BOS$^4$, which has an active volume of 320$\times$160$\times$80 mm$^3$. The BOS$^4$ detector was tested in a series of experiments using cosmic rays, protons, $\gamma$ rays and neutrons. Both simulations and experimental data indicate that a good $\it{n}$-$\gamma$ discrimination is achieved above 5 MeV$_{\rm ee}$ by means of the range difference of secondary particles induced by neutrons and $\gamma$ rays in the plastic scintillator. By employing the range discrimination, accidental (time-uncorrelated) $\gamma$ rays can be efficiently suppressed, thus enabling a lower detection threshold. This discrimination technique shows an advantage from other existing methods for plastic scintillator, which usually employs the conventional TOF method and suffers from accidental $\gamma$ rays. With the capability of high rate detection, the S$^4$ detector is superior in a high background environment, and can make an ideal neutron detector especially when neutron events are overwhelmed by $\gamma$-ray background. The detection efficiency of the detector was determined at two neutron energies by the $\it{d}$+$\it{d}\to\it{n}$+$^{3}$He reaction, and a good agreement was obtained between experimental data and Monte Carlo simulations using the commonly-used models in the Geant4 code.

\section{Future prospect}
\label{6}
As shown in Fig. \ref{SRNG}, there are always some neutrons distributed randomly between $\rho$ values of $-1$ and 1 and thus mis-identified as $\gamma$ rays, even at a very low threshold. One possible reason is that some of the conversions occur near the interface between layers and the low-energy secondary particles from neutrons can easily reach the next layer. In order to reduce such mis-identification of neutrons, especially for low-statistic experiments, we suggest to use three separate readouts rather than only two (odd and even) readouts, namely the output of one layer in every three layers are connected to one readout. Instead of the balance ratio $\rho$, a Dalitz plot can then be defined by the three readouts. Considering the bordering reactions, protons or carbon ions from neutrons below 100 MeV are expected to stop within two layers after conversions, whereas electrons from $\gamma$ rays with the same energy most likely penetrate three layers or more. By differentiating the number of penetrated layers up to three, the three-separate-readout method is expected to provide further versatility for the identification of neutrons.

\section*{Acknowledgement}
We thank the RCNP Ring Cyclotron staff for delivering the proton and deuteron beams stably throughout the experiments. H.J.Ong and D.T.Tran appreciate the support of Hirose and Nishimura International Scholarship Foundations, respectively. I.Tanihata acknowledge the support of the PR China government and Beihang University under the Thousand Talent Program. This work was supported in part by Grand-in-Aid for Scientific Research No. 23224008 from Monbukagakusho, Japan.




\section*{Reference}
\begin{thebibliography}{99}


\bibitem{p2pn}
A. Tang, et al., Phys. Rev. Lett. 90 (2003) 042301.
\bibitem{pnd}
D. Albrecht, et al., Nucl. Phys. A 322 (1979) 512.
\bibitem{eepn1}
D.G. Middleton, et al., Eur. Phys. J. A. 29 (2006) 261.
\bibitem{eepn2}
R. Subedi, et al., Science 320 (2008) 1476.
\bibitem{LAND}
Th. Blaich, et al., Nucl. Instrum. Methods Phys. Res. Sect. A 314 (1992) 136.
\bibitem{SubediPhD}
R. Subedi, Ph.D. thesis, Kent State University, 2007.
\bibitem{HANDPRL}
I. Korover, et al., Phys. Rev. Lett. 113 (2014) 022501.
\bibitem{PSD}
J.B. Birks, The Theory and Practice of Scintillation Counting, Pergamon Press, London, 1964.
\bibitem{liquid}
F.D. Brooks, Nucl. Instrum. Methods 162 (1979) 477.
\bibitem{DEMON}
I. Tilquin, et al., Nucl. Instrum. Methods Phys. Res. Sect. A 365 (1995) 446.
\bibitem{NeutronShell}
D.G. Sarantites, et al., Nucl. Instrum. Methods Phys. Res. Sect. A 530 (2004) 473.
\bibitem{EJ299}
S.A. Pozzi, et al., Nucl. Instrum. Methods Phys. Res. Sect. A 723 (2013) 19.
\bibitem{CSDA}
M.J. Berger, et al.(2005), ESTAR, PSTAR, and ASTAR: Computer Programs for Calculating Stopping-Power and Range Tables for Electrons, Protons, and Helium Ions (version 1.2.3). Retrieved February 7, 2017 from $\langle$http://physics.nist.gov/PhysRefData/Star/Text/ESTAR.html$\rangle$.
\bibitem{geant4}
S. Agostinelli, et al., Nucl. Instrum. Methods Phys. Res. Sect. A 506 (2003) 250. Official website: http://geant4.cern.ch/.
\bibitem{10.2}
Geant4 10.2 Release Notes, http://geant4.cern.ch/support/Release\\Notes4.10.2.html.
\bibitem{INCL1}
A. Boudard, et al., Phys. Rev. C 87 (2013) 014606.
\bibitem{INCL2}
D. Mancusi, et al., Phys. Rev. C 90 (2014) 054602.
\bibitem{MeVee}
K. Nakayama, E.F. Pessoa, R.A. Douglas, Nucl. Instrum. Methods 190 (1981) 555.
\bibitem{GRAF}
C. Iwamoto, et al., Commissioning of CAGRA + Grand Raiden experiment, RCNP Annual Report 2014.
\bibitem{GR}
M. Fujiwara, et al., Nucl. Instrum. Methods Phys. Res. Sect. A 422 (1999) 484.
\bibitem{Leo}
W.R. Leo, Techniques for Nuclear and Particle Physics Experiments: A How-to Approach, Springer-Verlag Berlin Heidelberg, New York, 1994.
\bibitem{BERT}
D.H. Wright and M.H. Kelsey, Nucl. Instrum. Methods Phys. Res. Sect. A 804 (2015) 175.
\end{thebibliography}


\end{document}